\documentclass[12pt,a4paper]{article}

\usepackage[british]{babel}

\usepackage[a4paper,top=1.5cm,bottom=1.75cm,left=1.75cm,right=1.75cm,marginparwidth=1.75cm]{geometry}

\usepackage[style=apa, backend=biber]{biblatex} 
\DeclareLanguageMapping{british}{british-apa} 
\DeclareFieldFormat[article]{volume}{\apanum{#1}} 

\usepackage{hyperref} 
\hypersetup{               
    colorlinks=true,                
    breaklinks=true,                
    urlcolor= black,                
    linkcolor= blue,                
    filecolor=black,
    citecolor=blue}
\usepackage{amsmath}
\usepackage{graphicx}
\usepackage{multirow}
\usepackage{xcolor}
\usepackage{tcolorbox}
\usepackage{svg}
\usepackage[title]{appendix}
\usepackage{mathrsfs}
\usepackage{amsfonts}
\usepackage{orcidlink}
\usepackage{tabularx}
\usepackage{booktabs} 
\usepackage{threeparttable} 
\usepackage{algorithm}
\usepackage{algorithmicx}
\usepackage{algpseudocode}
\usepackage{listings}
\usepackage{enumitem}
\usepackage{chngcntr}
\usepackage{booktabs}
\usepackage{lipsum}
\usepackage{subcaption}
\usepackage{authblk}
\usepackage[T1]{fontenc}    
\usepackage{csquotes}       
\usepackage{diagbox}
\usepackage{comment}
\usepackage{lmodern}
\usepackage{etoolbox} 
\usepackage{helvet}  
\setlist{noitemsep} 
\usepackage{setspace}
\usepackage[title]{appendix}
\usepackage{float}   
\usepackage{setspace}
\usepackage{titlesec}
\titleformat{\section} 
  {\normalfont\Large\bfseries}{\thesection.}{1em}{}



\title{Telephone Surveys Meet Conversational AI: Evaluating a LLM-Based Telephone Survey System at Scale}

\author[1,2]{Max M. Lang \orcidlink{0009-0004-6815-5321}}
\author[1]{Sol Eskenazi}

\affil[1]{\small 60 Decibels Inc., United States of America}
\affil[2]{\small University of Oxford, United Kingdom}
\date{}
\begin{document}
\maketitle
\thispagestyle{empty}
\definecolor{60db}{HTML}{FF8075}
\definecolor{boxbg}{HTML}{fafaf7}  
\begin{center}
    \begin{tcolorbox}[
        colframe=60db, colback=boxbg, arc=2mm, boxrule=1pt, width=\textwidth, sharp corners=south, fonttitle=\bfseries,
        ]
        
        \begin{flushright}
            \includegraphics[scale=0.05]{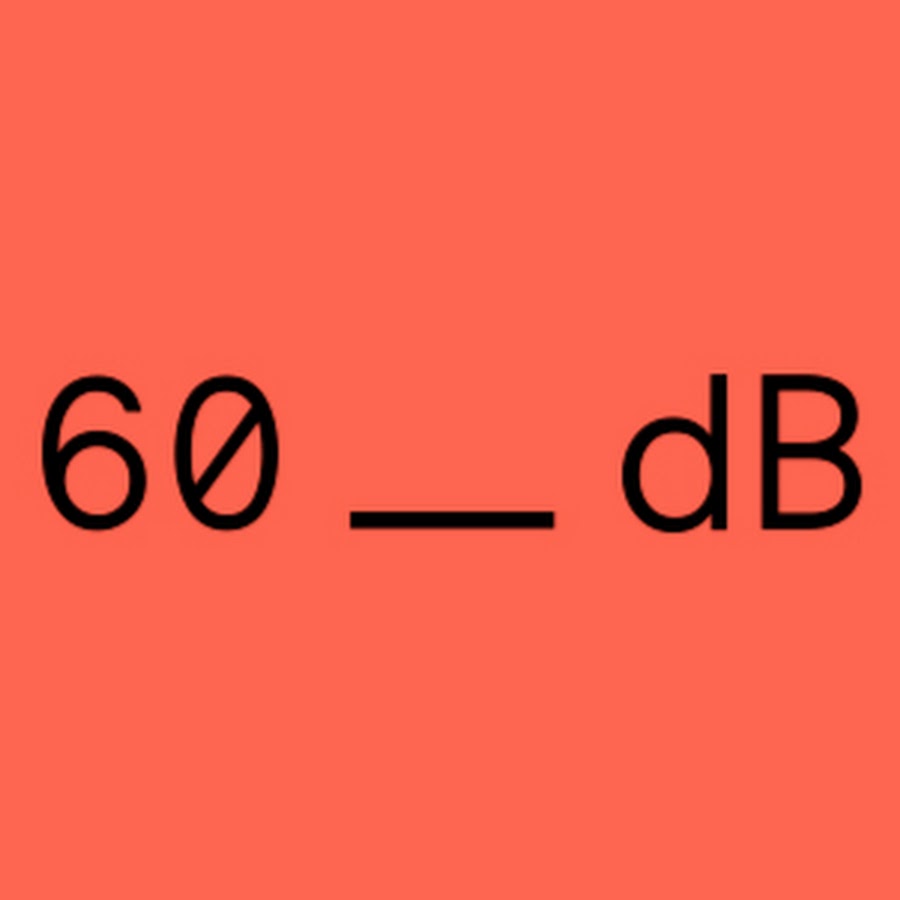} 
        \end{flushright}

        \textbf{\LARGE Abstract} \\[6pt]

        Telephone surveys remain a valuable tool for gathering insights but typically require substantial resources in training and coordinating human interviewers. This work presents an AI-driven telephone survey system integrating text-to-speech (TTS), a large language model (LLM), and speech-to-text (STT) that mimics the versatility of human-led interviews (full-duplex dialogues) at scale. 

        We tested the system across two populations—a pilot study in the United States (\(n=75\)) and a large-scale deployment in Peru (\(n=2,739\))—inviting participants via web-based links and contacting them via direct phone calls. The AI agent successfully administered open- and closed-ended questions handled basic clarifications, and dynamically navigated branching logic, allowing fast large-scale survey deployment without interviewer recruitment or training.

        Our findings demonstrate that while the AI system's probing for qualitative depth was more limited than human interviewers, overall data quality approached human-led standards for structured items. This study represents one of the first successful large-scale deployments of an LLM-based telephone interviewer in a real-world survey context. The AI-powered telephone survey system has the potential for expanding scalable, consistent data collecting across market research, social science, and public opinion studies, thus improving operational efficiency while maintaining appropriate data quality for research. 
        
        \vspace{10pt} 
        \textbf{Date:} \today \\
        \textbf{Website:} \href{https://www.60decibels.com}{\textcolor{60db}{https://www.60decibels.com}}

    \end{tcolorbox}
\end{center}

\newpage

\section{Introduction}
Telephone surveys have long been a cornerstone of data collection in social research, public opinion polling, and impact evaluation. Traditionally, they are conducted through either Computer-Assisted Telephone Interviewing (CATI) or Interactive Voice Response (IVR) systems. CATI allows human interviewers to follow a scripted questionnaire on a computer, enabling real-time adjustments and rapport-building. In contrast, IVR systems fully automate the process, using pre-recorded questions with respondents providing answers via keypad input or simple speech recognition. Efforts to automate survey interviewing began throughout the 1990s and early 2000s  with rule-based voice systems and early automated dialog systems \parencite{levin1998automatic,singh1999reinforcement, cole1997experiments, zeigler1994dialog,boyce2000natural,stent2006dialog}.  

In the 2010s, \cite{johnston2013spoken} conducted one of the first large-scale evaluations of an automated spoken dialog system administering social surveys, finding that with appropriate dialog management (e.g., response confirmation), such systems could approximate data quality levels seen in CATI. The study was part of a larger research project comparing different modes of survey administration \parencite{conrad2013mode}, motivated by \cite{conrad2007envisioning} in the book \textit{
Envisioning the Survey Interview of the Future} which remains relevant until today. More recent advancements in AI-driven interviewing, such as \cite{devault2014simsensei}, showed that virtual AI interviewers could elicit more candid responses than human interviewers, particularly in sensitive topics, due to reduced social desirability bias. 
These systems paved the way for more advanced automated survey methods. Recent research has leveraged the rapid advancements in speech-to-text (STT) models, large language models (LLMs), and text-to-speech (TTS) engines, significantly enhancing the naturalness and adaptability of automated interviewers \parencite{inoue2020job, nagasawa2023adaptive, ge2022should, zeng2023question, cuevas_collecting_2024}.  \cite{zeng2023question} demonstrated that LLMs could conduct semi-structured interviews, and \cite{wuttke2024ai} showed that LLMs were able to conduct conversational interviews, retrieving data comparable to traditional methods with additional scalability. Taking into consideration studies from the domain of medical interviewing, \cite{wang2024telephone} and \cite{hong2022} both conducted small-scale assessments of automated telephone follow-ups or web-based interviews when prior contact has been established.

Clearly automated interviewing has emerged as an interdisciplinary field of survey methodology, natural language processing, and human-computer interaction. Recent studies within the domain of survey research do employ LLMs but tend to focus on primarily text-based interactions in smaller samples or to support semi-structured follow-up generation rather than conducting a full interview flow \parencite{ge2022should, zeng2023question, wuttke2024ai}. 
\newline
We present a large-scale telephone survey system that seamlessly integrates STT, a transformer-based LLM, and TTS to fully mimic the flexibility of a human enumerator—i.e., one that can ask open-ended questions, interpret responses, and spontaneously formulate subsequent queries without strictly following a preset script. This set of capabilities has been described as full-duplex Spoken Dialogue Models (SDMs) in general literature \parencite{lin2022duplex, wang2024full, zhang2025llm}.

We aim to bridge the gap between the extensive literature of telephone surveys, automated interviewing systems, and conversational AI by developing and deploying an automated telephone survey system that integrates a pipeline of STT $\leftrightarrow$ LLM $\leftrightarrow$ TTS at scale. In collaboration with 60 Decibels Inc., a global impact measurement firm, we deployed a scalable AI interviewer that uses state-of-the-art conversational AI to carry out phone interviews on a large scale. This work advances prior research by demonstrating a fully automated, scalable telephone survey system that uses the recent advancements in conversational AI technology to emulate human enumerators at scale. Such a system has broad relevance: while motivated by impact evaluation, a robust AI telephone interviewer could be applied to social science research, opinion polling, and customer feedback surveys, especially in regions or populations where telephone outreach remain crucial.

\section{Methods}
This section outlines the development and implementation of the AI survey system, including its architecture, deployment strategies, and evaluation methods. We first describe the system’s core components and their role in enabling a human-like telephone-based survey. We then detail survey deployment and introduce our study population, including recruitment strategies. Finally, we present our approach to evaluating the performance of the automated telephone-based interviewing system.

\subsection{AI Survey System Architecture}
\begin{figure}[ht]
    \centering
    \includegraphics[width=0.95\linewidth]{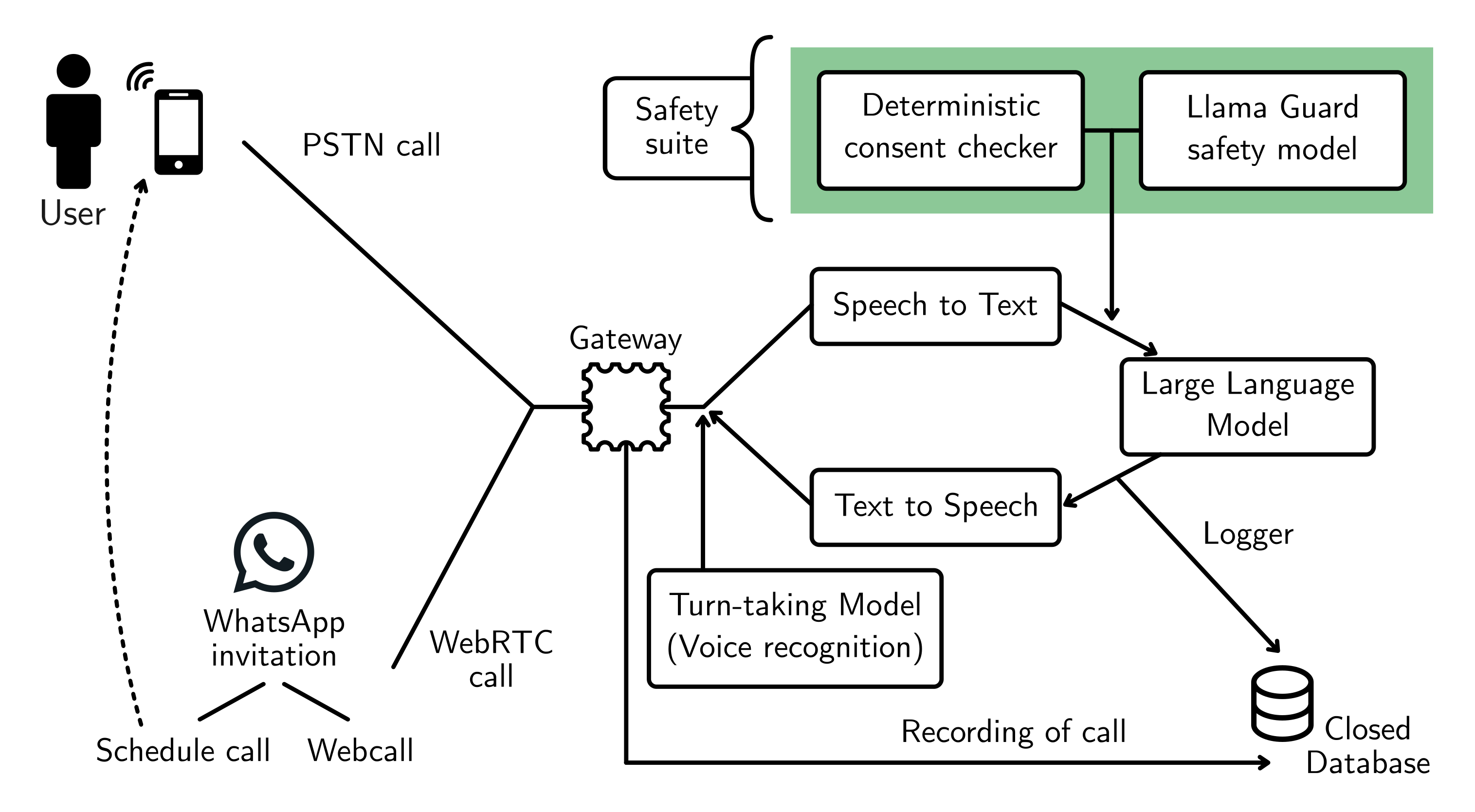}
    \caption{System architecture of the voice-based conversational AI survey agent}
    \label{fig:system-architecture}
\end{figure}

The survey system was implemented as a voice-based conversational AI agent that conducts phone surveys in natural language, mimicking a human enumerator (see \autoref{fig:system-architecture}). The pipeline integrates three key components—STT $\leftrightarrow$ LLM $\leftrightarrow$ TTS. First, an automated speech recognition module transcribes participants’ spoken responses into text (STT), then the transcription gets passed into an LLM that generates the next appropriate reply based on the conversation history, and finally, a TTS engine converts the LLM’s textual output into a natural-sounding voice response. These components worked in sequence to enable a real-time dialog: participants heard the AI agent’s questions, responded orally, and the system replied, creating an interactive interview experience (see \autoref{fig:system-architecture}). Alongside these three components, we implemented a turn-taking model that allows the user to interrupt the AI after a set number of (user) spoken words, triggers idle messages if the user remains silent, and defines a silence timeout to manage pauses effectively. 

All models used for each stage were internally fine-tuned versions of state-of-the-art LLMs, optimized for this survey domain. Specific model identities are omitted due to confidentiality, but each was internally evaluated to perform with high transcription accuracy and coherent, contextually appropriate dialog generation. All processing was performed in real time to maintain a smooth conversational flow. We also tested non-fine-tuned models to assess their general suitability for survey administration and found that they can function adequately for broad topics. However, given the specialized nature of application (impact research) and our existing data resources, we opted to employ a fine-tuned approach to achieve more precise handling of domain-specific questions and to maximize the performance of the system.

\subsection{Survey Deployment and Sampling}
We conducted a pilot study in the United States followed by the main study in Peru, with different sample characteristics and recruitment procedures for each. In the U.S. pilot, we recruited $n = 75$ participants from a general population sample as part of an impact research survey. In Peru, we targeted $n = 2,739$ participants who were primarily university students on behalf of a client that initiated the survey and recruited the participants. The pilot study was conducted in October 2024 and the main study was conducted over the first two weeks of November 2024.

We employed three outreach strategies to connect participants with the AI survey agent. We used WhatsApp invitations that included links for both a (1) web-based call and (2) call scheduling, and conducted (3) direct telephone calls made during pre-determined time windows.

\begin{figure}
    \centering
    \includegraphics[width=0.86\linewidth]{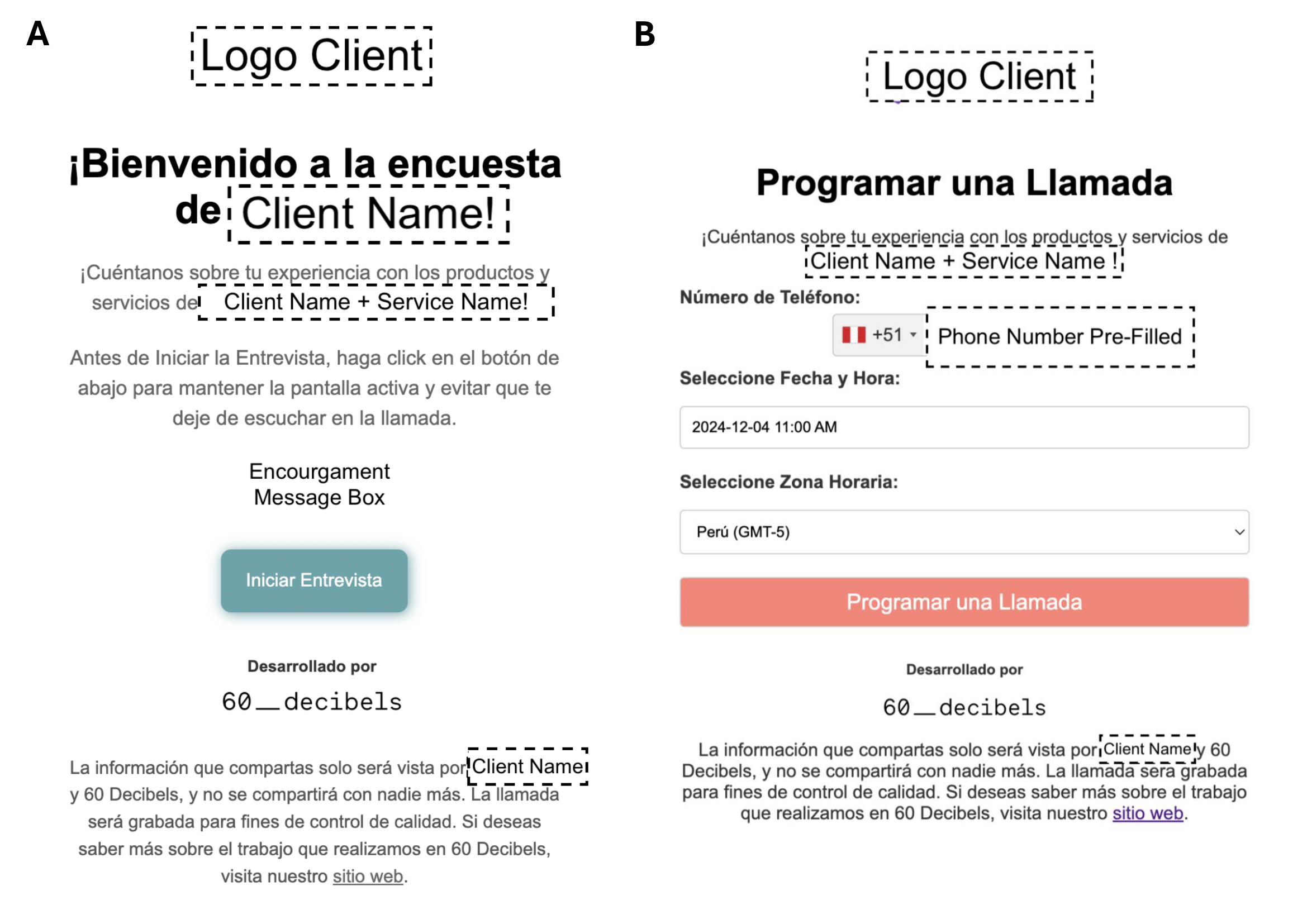}
    \caption{Web Calling and Call Scheduling Interfaces: Panel A shows the Web Calling interface, where participants start an AI survey call via WebRTC directly in their browser. Panel B displays the Call Scheduling interface, allowing users to schedule a call with a pre-filled phone number and time zone.}
    \label{fig:UI}
\end{figure}

\begin{enumerate}
    \item \textbf{WhatsApp Invitations} \newline
    Participants received a WhatsApp message that included a brief introduction to the survey mentioning that they would be talking to an AI and two personalized links: one to initiate a web-based call and another to schedule a call at a later time. By clicking the first link, participants accessed an in-browser calling interface with on-screen survey instructions (see \autoref{fig:UI}). This interface automatically connected them to an AI agent via a WebRTC call \parencite{uberti_webrtc_2024}. As users engaged with the survey, they interacted with the browser window, speaking towards it. To enhance engagement, they received encouragement messages at 25\%, 50\%, and 75\% completion.
    The second link directed participants to a scheduling interface. Their phone number and time zone were pre-filled, although both could be updated if necessary. Users then selected a preferred date and time window, upon which the system automatically initiated a call via the AI agent at the chosen time. 
    \item \textbf{Direct Telephone Calls}\newline
    We conducted direct telephone calls to participants during pre-determined "optimal"  time windows based on our experience from previous surveys in the region.
\end{enumerate}

All U.S. participants received the personalized links (web call or option to schedule a call) through WhatsApp, and no direct phone calls were made for the pilot study. The pilot primarily served to test the system’s performance and refine the survey flow. As an incentive each U.S. participant who completed the survey was offered a \$10 Amazon gift card.

For the Peruvian survey, we sent out $200$ WhatsApp invitations (WebRTC call link, and the scheduling option) and conducted $2,539$ direct calls via a local Peruvian number for caller ID \parencite{uberti_webrtc_2024}. Peruvian calls and messages were conducted in Spanish with a Peruvian accent. As an anti-fraud and engagement measure, participants who received direct calls were sent a primer via SMS a day in advance to inform the person that they would receive a call by the number used for the survey to give feedback on the client's service. To encourage participation, those who completed the survey in Peru were entered into a raffle to win a pair of headphones, which was organized by the client who initiated the survey. Participants who missed the call received a voicemail introducing the study and inviting them to redial the same local number to initiate the survey when they were available.

If a participant engaged with the AI agent via a web call or by answering the phone, the participants were greeted by the AI survey agent with their first name, which was obtained from a personalized hyperlink unique identifier or via the phone call setup. The agent then delivered a structured introductory message, stating its identity as a 60 Decibels AI virtual researcher, the purpose of the survey, the expected duration, and a prompt for consent. \autoref{fig:bilingual_greeting} shows the introduction used.
\begin{figure}[htb]
\begin{tcolorbox}[colback=lightgray!20, colframe=gray!50, sharp corners=southwest, boxrule=0.5pt]
    \textit{Spanish}: Hola \textbf{First Name}, soy \textbf{Name of AI Assistant}, un investigador virtual de 60 Decibels AI. Estoy realizando una investigación en nombre de \textbf{Client Name} con sus clientes de \textbf{Service Name} para conocerlos y obtener su opinión. ¿Tienes quince minutos para hablar conmigo?
    
    \vspace{0.5em} 
    
    \textit{English}: Hi \textbf{First Name}, I’m \textbf{Name of AI Assistant}. A 60 Decibels AI virtual researcher. I’m conducting research on behalf of \textbf{Client Name} with their \textbf{Service Name} customers to get to know them and get their feedback. Do you have fifteen minutes to talk to me?
    
\end{tcolorbox}
\caption{AI Assistant Introduction:  Original (Spanish) and translated (English) version of the Survey Participant Greeting}
    \label{fig:bilingual_greeting}
\end{figure}

Following the initial greeting, participants were informed about how their data would be used and with whom it would be shared. Both this disclosure and the consent prompt were managed through a deterministic approach, following guidelines laid out in existing literature \parencite{jiang2021supporting, bach2024unpacking}. If the participant’s response was any form of \textit{“No”} (declining consent), the system immediately terminated the survey call without any pressure or additional questions. The agent would acknowledge the decision (e.g. \textit{“Thank you for your time, have a great day”}) and end the call.

We also implemented safeguards to maintain appropriate and secure interactions throughout the conversation. All user inputs were monitored by a secondary LLM-based safety model, \textit{LlamaGuard3}, which acted as a real-time filter for any inappropriate content or attempt at prompt injection \parencite{dubey2024llama}. If a user’s input was flagged (for example, containing offensive language or instructions aiming to derail the survey), the system would intervene by either steering the conversation back on track or gracefully ending the interaction if needed.

The survey questionnaire consisted of closed-ended and open-ended items to collect both quantitative data and qualitative feedback. It included simple yes/no questions, NPS items with a range from 0 to 10, several Likert-scale questions to measure attitudes or levels of agreement, and a number of open-ended questions that invited participants to elaborate on their opinions or experiences in their own words \parencite{likert1932technique, reichheld2003the}. The questionnaire also included conditional branching logic for follow-ups: certain questions would only be asked based on a previous response or different questions would be asked based on the rating given. For example, if a participant answered \textit{“Yes”} to a yes/no question about having used a particular service the AI agent would ask an open-ended follow-up such as \textit{“Could you briefly describe your experience with it?”} Conversely, if the answer was \textit{“No”} the agent would skip that follow-up and move to the next topic. The overall survey length was designed to be about 10–15 minutes, though the actual duration varied depending on how much detail participants provided in open-ended responses. The survey questionnaire cannot be shared, as it contains proprietary information.

\subsection{Evaluation of the AI Survey System}

After the data collection phase, we conducted both quantitative and qualitative assessments to evaluate the AI agent’s performance in conducting a survey with real participants, the effectiveness of our outreach methods, and the quality of the responses elicited. We defined a fully completed interview (100\% survey completion) as a “successful response”, while interviews falling between 76\% and 100\% completion were categorized as “partially completed”. Response rates for both definitions were reported according to the outreach method. Both fully completed and partially completed interviews were considered in response rate calculations, following AAPOR’s Response Rate 1 (RR1) and Response Rate 2 (RR2) definitions \parencite{definitions2011final}. 

We documented the total call time for fully completed responses, measured from the moment the participant answered until the call ended. As for web calls participants could start multiple calls, we would select the call with the longest duration. Call outcomes were categorized based on participants’ progress through the survey transcript. If there was no answer, the outcome was recorded as "Not Picked Up" or for web calls "Not Clicked Through". For those who answered but progressed through less than 10\% of the survey, we noted whether they ended the call after learning it was an AI agent, or explicitly refused participation. Beyond this point, calls were considered according to the following progress intervals: 11–25\%, 26–50\%, 51–75\%, and 76–100\%.
\newline
To assess dialog dynamics, we calculated the total number of turns, defined as the sum of individual speaker utterances (either respondent or AI), User-AI ratio defined as $ \frac{\text{Number of User Responses}}{\text{Number of AI Responses}}$ and calculated the Flesch Reading Ease scores on the transcripts of the fully completed responses \parencite{flesch1948new}.
\noindent
We also analyzed the transcript content for both the AI agent and human respondents on a per-interview basis, however, only for fully completed interviews. We measured for both human respondents and AI agent:

\begin{itemize}
    \item Total number of sentences
    \item Total number of questions asked
    \item Average words per turn
\end{itemize}

\noindent
Beyond these quantitative metrics, we carried out qualitative reviews of the AI-collected fully completed responses to verify that their depth, consistency, and contextual alignment matched the standard attained by human enumerators. Specifically, our team evaluated whether the AI agent was able to \textbf{(i)} identify responses out of the allowed answer range (e.g., 1-10), and \textbf{(ii)} able to gather sufficiently detailed information for meaningful interpretation. Transcripts were then compared to data typically gathered by experienced human enumerators.

\section{Results}
This section presents the results of our study, beginning with response rates across different outreach methods in both the U.S. pilot and the main survey in Peru. We then analyze the structure and dynamics of AI-mediated survey interactions using quantitative metrics, including conversation length, turn-taking balance, and linguistic complexity. Finally, we provide a qualitative assessment of the AI's ability to engage participants, maintain coherence, and elicit meaningful responses in comparison to human interviewers.
\begin{figure}[htbp]
    \centering
    \includegraphics[width=0.95\linewidth]{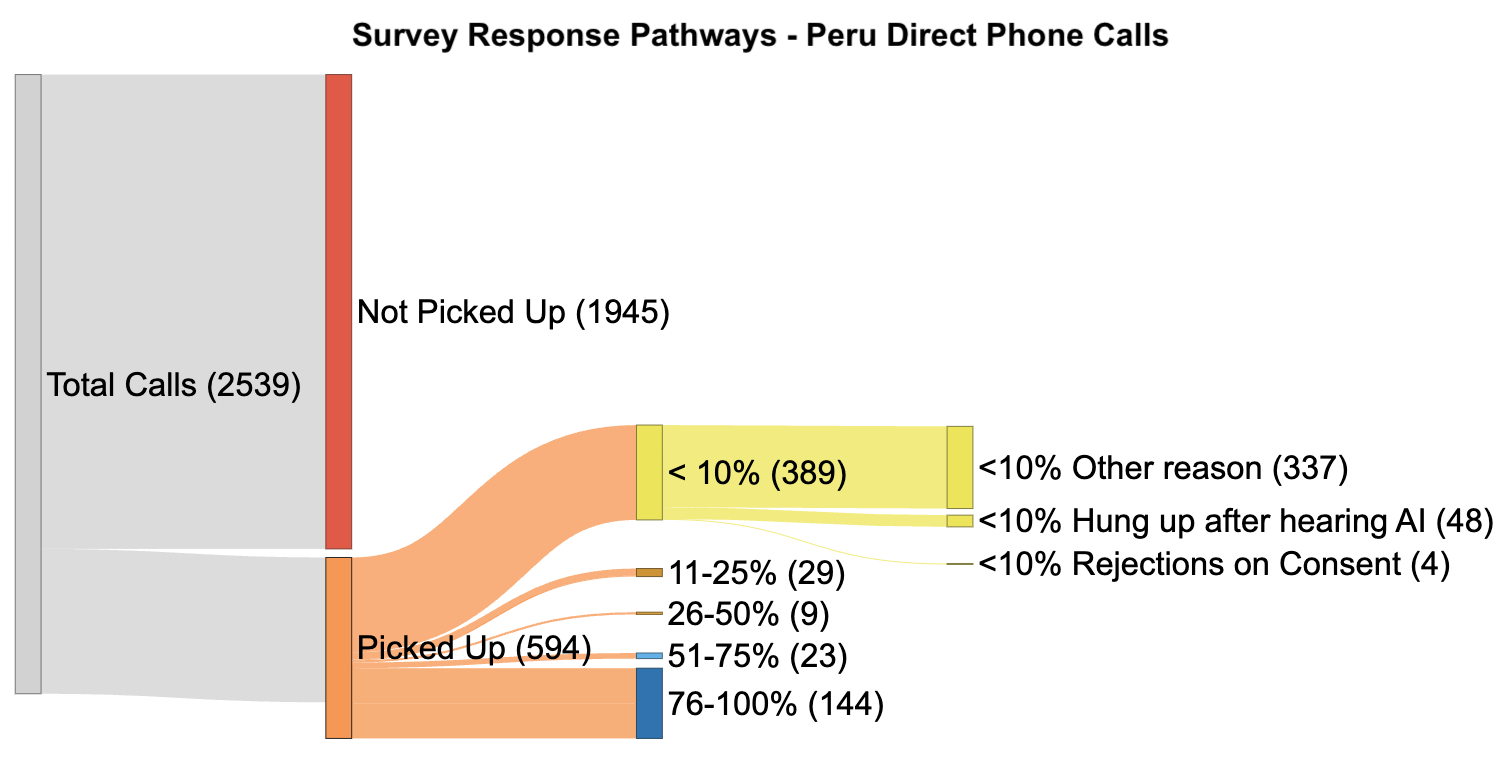}
    \caption{Sankey diagram illustrating the flow of phone call attempts in Peru.}
    \label{fig:phone_call_flow}
\end{figure}
\begin{figure}[htbp]
    \centering
    \includegraphics[width=0.95\linewidth]{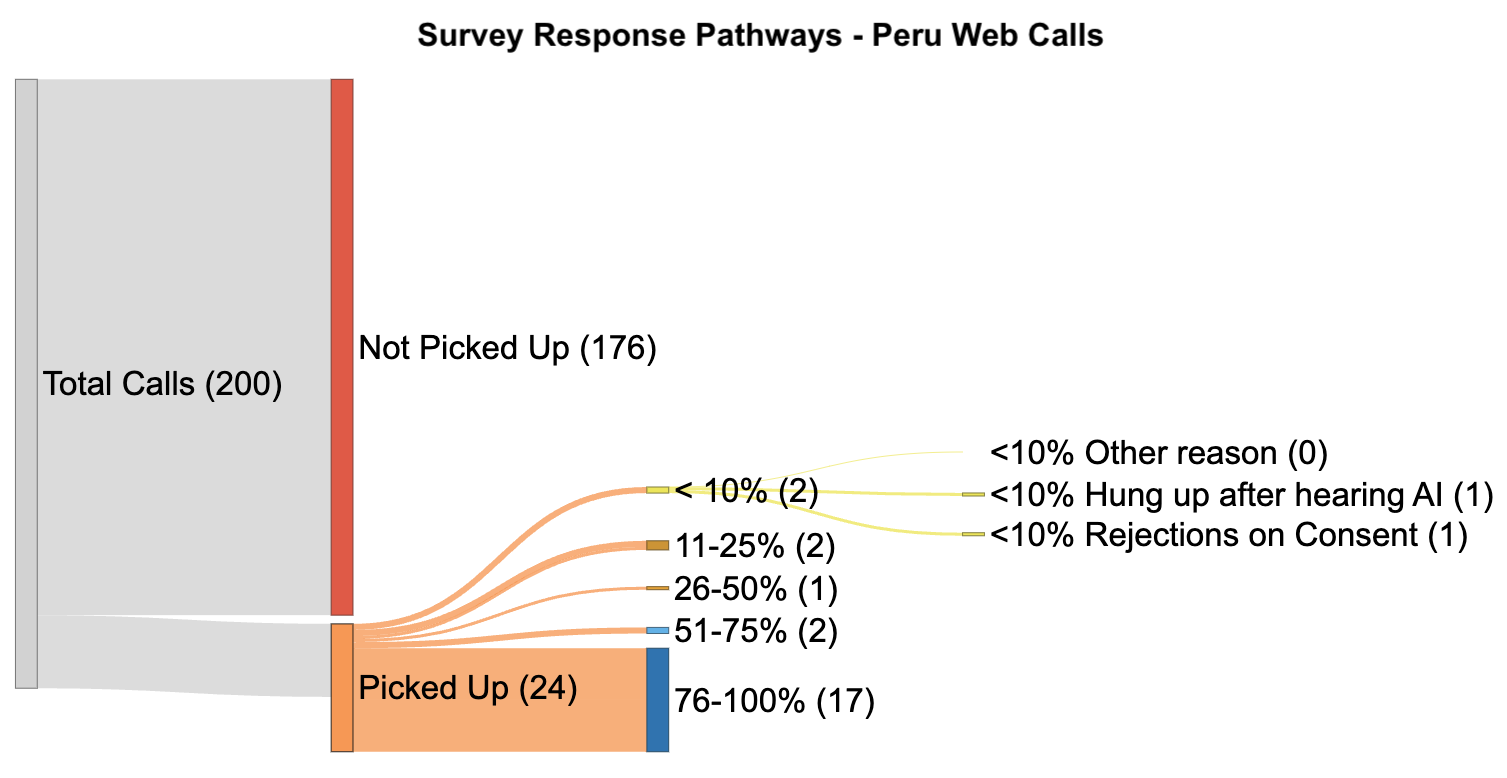}
    \caption{Sankey diagram illustrating the flow of web call attempts in Peru.}
    \label{fig:web_call_flow}
\end{figure}

\subsection{Response Rates}

In the U.S. pilot study, we observed an RR1 of 4\%. Out of 75 calls, 3 resulted in a fully completed survey (100\% of questions answered), and an additional 5 calls reached at least 75\% survey completion (RR2 6.7\%). Notably, no participants requested a scheduled call.

In the main study in Peru, we received 11 fully completed surveys (100\% completion) and 17 surveys with more than 75\% completion out of 200 web calls, corresponding to an RR1 of 5.5\% and RR2 of 8.5\%, respectively. For direct telephone calls (i.e., outbound calls initiated by the system), $131$ out of $2,539$ call attempts successfully reached a respondent and yielded fully completed surveys, while 144 calls resulted in a partially completed survey (>75\% of questions answered). We had three individuals calling the AI agent back after initially not picking up and then fully completing the survey, such that they are included in the 131 fully complete responses. These outcomes translated to an RR1 of 5.2\% and an RR2 of 5.7\%. As in the U.S. pilot, no participants in Peru requested to schedule a call.

\autoref{tab:comparison_table} summarizes and compares the results from the U.S. pilot and the Peru main study. 
\autoref{fig:phone_call_flow} and \autoref{fig:web_call_flow} illustrate the recruitment flow for phone calls and web calls in Peru using Sankey diagrams, showing how many calls connected to a respondent and ultimately led to a completed or partially completed survey.
In \autoref{sec-quant-an} and \autoref{sec-qual-an} we will only focus on fully completed interviews.
\begin{table}[ht]
    \caption{Comparison of Outreach Results in U.S. (Pilot) and Peru (Main)}
    \centering
    \renewcommand{\arraystretch}{1.0}
    \setlength{\tabcolsep}{5pt} 
    \begin{tabularx}{\textwidth}{cXXXcc}
        \toprule
        \multicolumn{6}{c}{\textbf{U.S. (Pilot)}} \\
        \midrule
        n & Outreach Method & Fully Completed & Partially Completed & RR1 & RR2 \\
        \midrule
        \multirow{2}{*}{75} & Webcall & 3 & 5 & \multirow{2}{*}{4\%} & \multirow{2}{*}{6.7\%} \\
        & Scheduled Call & 0 & 0 & & \\ 
        \midrule
        0 & Direct Call & - & - & - & - \\
        \midrule
        \multicolumn{6}{c}{\textbf{Peru (Main)}} \\
        \midrule
        n & Outreach Method & Fully Completed & Partially Completed & RR1 & RR2 \\
        \midrule
        \multirow{2}{*}{200} & Webcall & 11 & 17 & \multirow{2}{*}{5.5\%} & \multirow{2}{*}{8.5\%} \\
        & Scheduled Call & 0 & 0 & & \\ 
        \midrule
        2539 & Direct Call & 131 & 144 & 5.2\% & 5.7\% \\
        \bottomrule
    \end{tabularx}
    \label{tab:comparison_table}
\end{table}

\subsection{Quantitative Text Analysis}
\label{sec-quant-an}
As shown in \autoref{tab:conversation-metrics}, each conversation between the AI interviewer and a participant consisted of a series of turns, where a turn represents a single exchange by either the AI or the user. On average, each conversation consisted of 52.95 turns (median = 53) and lasted on average approximately 7:00 minutes (median = 6:38 minutes). The balance of interaction was nearly even, with a mean User-AI turn ratio of 0.96 (median = 0.96). The overall interview text had on average a Flesch Reading Ease score of 28.87 (median 26.13).

\begin{table}[ht]
\centering
\caption{Summary of Conversation Metrics with Minimum, Maximum, and Quartiles}
\label{tab:conversation-metrics}
\begin{tabular}{lrrrrrr}
\toprule
\textbf{Metric} & \textbf{Min} & \textbf{Q1} & \textbf{Median} & \textbf{Mean} & \textbf{Q3} & \textbf{Max} \\
\midrule
\multicolumn{7}{l}{\textbf{Overall Conversation}} \\
\midrule
Total turns per conversation                   & 39  & 49   & 53   & 52.95  & 57  & 91  \\
Duration of conversation                       & 4:21 & 5:53 & 6:38 & 7:00   & 7:29 & 12:43 \\
User-AI turn ratio                          & 0.92 & 0.96 & 0.96 & 0.96   & 0.97 & 1.00 \\
Overall Flesch Reading Ease                    & 10.77 & 23.92 & 26.13 & 28.87  & 30.85 & 45.95 \\
\midrule
\multicolumn{7}{l}{\textbf{AI Agent}} \\
\midrule
Number of AI conversational turns              & 20  & 25  & 27  & 26.94 & 29  & 46  \\
Total AI sentences                             & 31  & 45  & 50  & 51.29 & 56  & 88  \\
Number of AI questions                         & 13  & 17  & 19  & 19.14 & 21  & 30  \\
Words per AI turn                              & 17.00  & 21.82 & 23.00 & 23.35 & 24.59 & 34.64 \\
\midrule
\multicolumn{7}{l}{\textbf{Participant}} \\
\midrule
Number of participant conversational turns     & 19  & 24  & 26  & 26.01 & 28  & 45  \\
Total participant sentences                    & 11  & 19  & 22  & 22.92 & 26  & 42  \\
Number of participant questions                & 0   & 1   & 1   & 2.05  & 3   & 12  \\
Words per participant turn (overall)           & 1.47 & 2.89 & 4.20 & 5.59  & 6.22 & 31.51 \\
Words per participant turn (open-ended)        & 3.33 & 9.00 & 13.00 & 18.58 & 22.38 & 65.50 \\
\bottomrule
\end{tabular}
\end{table}

Next, we examined the content of both the AI agent’s prompts and the participants’ responses. 
The AI agent’s survey script contained on average 26.94 conversational turns (median 27) and on average 51.29 total sentences (median 50) spoken by the AI in each conversation. Within these, the AI asked on average 19.14 questions (median 19) per interview. 
Human participants contributed an average of 26.01 conversational turns (median 26) and responded on average 22.92 total sentences (median 22). The participants themselves asked few questions, with an average of 2.05 questions per interview (median 1).

Each AI conversational turn contained on average 23.35 words (median 23). Human utterances were typically shorter, at a mean of 5.59 words per turn (median 4.20). However, this includes brief answers to yes/no questions and Net Promoter Score (NPS) ratings, where participants might respond with a single number. For open-ended questions, participants contributed on average 18.58 words per response (median 13).

\begin{figure}[htbp]
    \centering
    \includegraphics[width=0.95\linewidth]{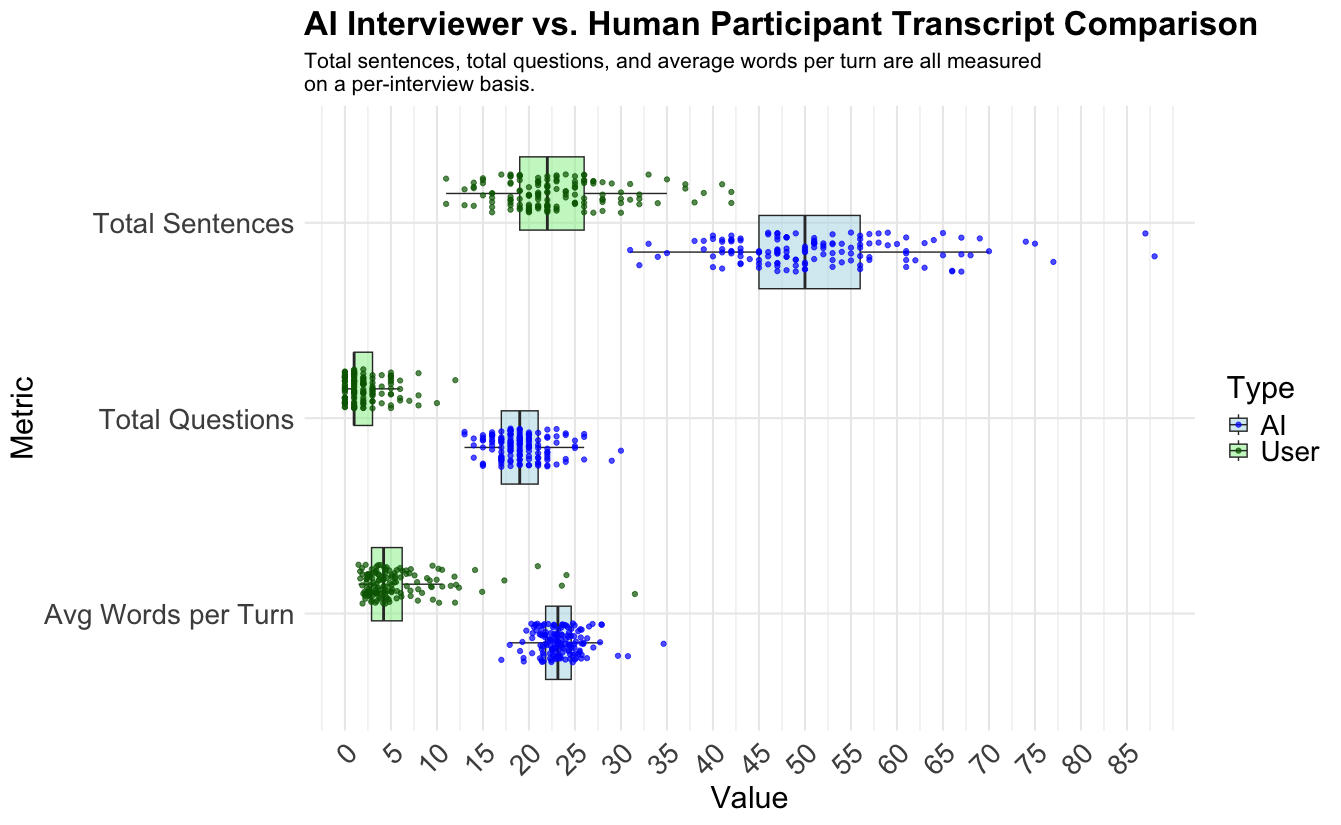}
    \caption{Comparison of AI vs.\ Human Participant Transcript Metrics (total sentences, total questions, and average words per turn).}
    \label{fig:transcriptanalysis}
\end{figure}

\autoref{fig:transcriptanalysis} illustrates the per interview metrics for both AI and human utterances (e.g., total sentences, total questions, and average words per turn) and in the appendix \autoref{fig:call_duration} displays the duration distribution of fully completed interviews, most of which range roughly from 5 to 9 minutes. In the boxplots for “Total Sentences” and “Total Questions,” each dot represents a single interview, whereas in the boxplot for “Average Words per Turn,” each dot reflects the average words per turn for that interview.

\subsection{Qualitative Text Analysis}
\label{sec-qual-an}

In addition to the quantitative metrics, we examined the call transcripts qualitatively. In terms of consistency and contextual alignment, the AI agent generally succeeded in detecting contradictory responses or answers that fell outside the scope of the question, allowing it to guide most participants effectively through the entire survey. The majority of respondents provided suitable answers on the first attempt, particularly for categorical questions. When faced with ambiguity or incompleteness, the AI agent followed up to clarify and prompt respondents toward an appropriate response category. The AI agent's performance was comparable to that of an experienced human enumerator.

Regarding depth, the AI agent did not reach the human benchmark in probing for richer information, especially during open-ended questions or when participants seemed hesitant to elaborate. As a result, responses were often shorter and less rich than those elicited by human interviewers, yet still exceeded the level of detail typical of online self-administered surveys.

\section{Discussion}
This study provides one of the first large-scale demonstrations of automating telephone surveys using an integration of an LLM with TTS and STT technologies. We deployed an AI “enumerator” capable of conducting entire phone interviews without human assistance. Contributing to the existing literature on phone surveys and automated voice surveys, we show that the current state of conversational AI can provide a user experience that goes beyond simple interactive Voice Response systems and more closely mimics a human enumerator \parencite{conrad2013mode, devault2014simsensei, wuttke2024ai, johnston2013spoken}. Moreover, our work contributes to the evolving literature of full-duplex Spoken Dialogue Modeling and its applications \parencite{lin2022duplex, wang2024full, zhang2025llm, jin2021duplex, lu2025duplexmamba}.  Finally, our work shows that a fully AI-driven telephone survey can be effectively scaled for real-world data collection.

The AI agent successfully completed hundreds of interviews following a non-trivial questionnaire which included several conditional branches and did not expose any major hallucinations, such as making up questions that were not in the questionnaire. The implementation of a turn-taking model and idle messages was beneficial, as it prevented silence timeouts for persons eager to continue, despite network issues preventing the transcription from reaching the LLM stage of our system. Interestingly, the AI survey agent was able to ensure that respondents gave valid and complete answers. It could detect when a response was unclear or outside expected parameters (for example, if a numeric answer was out of range) and would politely prompt the respondent to correct or clarify, much as a human interviewer would. This helped in obtaining usable data and in preventing respondent mistakes. However, we found that the AI agent lacked the probing ability that human interviewers have for eliciting richer qualitative insights, which can be seen in the lower average duration (7:00 minutes) of a questionnaire that humanly led would take approximately 15 minutes. If a respondent’s answer was brief or vaguely worded, the AI typically accepted it and moved on, whereas a human interviewer might have asked \textit{“Could you tell me more about that?”} or rephrased the question to probe deeper. This is consistent with recent literature and motivates the use of a potential multi-agent system that uses a separate model for probing and generating follow-up questions \parencite{wuttke2024ai, zeng2023question, ge2022should, wei2024leveraging, cuevas_collecting_2024}. We are hypothesizing that the limited probing ability might be due to implemented safety measures of foundational models, as the model rejects to fixate on, for example, a personal problem of a human participant \parencite{abdulhai2023moral, ranaldi2023large, rottger2023xstest}. Nevertheless, we found that the open-ended responses were still of better quality than responses received through online self-administered surveys.
\newline
Our deployment included both web-based voice calls and standard telephone calls, and we observed notable differences in response quality between these modes. Web-based calls often suffer from competing for the participant's attention, as respondents might have been distracted by on-screen notifications or other apps. As a result, their answers during web calls seemed often rushed or less thoughtful. In contrast, when the AI reached participants via a traditional phone call, respondents appeared more focused on the conversation. Although scheduling a traditional phone call—rather than sending a link for a web call—gave the research team control over the exact timing, this approach did not always align with the unknown schedules of potential respondents. Given that the AI agent would leave a voicemail and is available around the clock, this offers a best-of-both-worlds solution, as participants could receive or return survey calls at any time. We found that the three respondents who opted to call back were committed to providing rich feedback and valuable insights, which otherwise might have been missed. Moreover, we discovered that this method facilitates fast, near real-time deployment of telephone surveys at scale, since it eliminates the requirement for training and practice time required for human enumerators. However, this comes with potential trade-offs in data quality that researchers must consider depending on their objectives.

\subsection{Limitations}
This research was conducted in an industry setting, which influenced certain methodological choices. Notably, we did not incorporate specific study designs, such as a human enumerator control group for the questionnaire, as all data in this campaign were collected exclusively by the AI agent. Consequently, this approach limits our ability to derive deeper insights into the system’s comparative performance. Additionally, several factors limit the generalizability of these results. First, the sample of the main study in Peru primarily consisted of university students, who may be more technically savvy and more open to AI-driven interactions than the general population \parencite{nikolenko2023attitude, horowitz2024adopting}. As a result, the findings should be interpreted with caution when discussing the potential challenges or varying levels of comfort in broader demographic groups. Second, while the AI system functioned autonomously during the interviews, human intervention was still necessary for tasks such as scheduling, sending reminders, and monitoring the system’s performance. Consequently, a fully automated pipeline from participant recruitment to data collection remains an objective for future research. Third, we only analyzed completed interviews and did not receive feedback from those who dropped out, limiting our understanding of drop-off reasons and overall user experience. Finally, for confidentiality reasons, we do not disclose the specific (fine-tuned) models used in the system, which may limit the reproducibility of our approach in other contexts.

\subsection{Future Research}
Future research should expand participant pools beyond university students to validate the generalizability of these findings among older adults, and other demographic groups. Since LLM-based systems can be adapted for multiple languages, testing the AI enumerator in different linguistic contexts would be an important step to determine whether certain languages or dialects pose unique challenges for automated survey interviewing using state-of-the-art TTS and STT. Efforts to improve the AI system’s probing and contextual understanding are also warranted, potentially through dynamic scripts or multi-agent approaches for generating follow-up questions. Running a similar campaign with human enumerators as a baseline would contribute to a better understanding of the AI agent’s capabilities for phone surveys. Finally, further exploration of TTS options, including different voice types and degrees of expressiveness, may reveal strategies to increase engagement and reduce attrition, as traditional phone surveys found a difference in response behavior based on the interviewer's voice \parencite{irvine2013not, van2004event,groves2008telephone, schober2015precision}. By addressing these aspects, it may be possible to achieve a fully automated, highly adaptable telephone survey system, balancing efficient data collection with the depth and quality of respondent feedback traditionally associated with human-led interviews.

\section{Conclusion}
This study presents and evaluates a fully automated, AI-driven telephone survey system integrating STT, LLM, and TTS for large-scale data collection. Our findings demonstrate that this approach can reliably engage respondents in real-time, automatically prompt clarifications, and capture both quantitative and open-ended responses. However, the AI agent did not probe for deeper, more detailed elaborations in open-ended questions as thoroughly as human interviewers typically do. Moreover, additional human oversight is currently required for tasks like scheduling calls or sending reminders and monitoring system performance, limiting the true end-to-end automation of the pipeline. Despite these challenges, our results underscore the practical viability of leveraging conversational AI for phone-based research at scale. Future efforts could focus on improving the AI’s probing strategies, expanding language coverage, and comparing AI-driven interviews to human-led surveys in diverse populations. With continued refinements, this system has the potential to serve as a scalable and rapidly deployable alternative to traditional telephone surveying methods.

\section*{Acknowledgments}
We thank 60 Decibels for supporting this research, with special appreciation to Sasha Dichter, Adriana Baqueiro, Ramiro Rejas, and the entire LATAM team for their contributions. 
\noindent
MML also acknowledges Frauke Kreuter for recognizing the potential of this project early on, providing continuous support, and facilitating the connection with 60 Decibels.   
\noindent
Thank you for making this research possible.

\printbibliography

@inproceedings{stent2006dialog,
  title={{Dialog systems for surveys: The Rate-a-Course system}},
  author={Stent, Amanda and Stenchikova, Svetlana and Marge, Matthew},
  booktitle={{2006 IEEE Spoken Language Technology Workshop}},
  pages={210--213},
  year={2006},
  organization={IEEE}
}

@inproceedings{johnston2013spoken,
  title={{Spoken dialog systems for automated survey interviewing}},
  author={Johnston, Michael and Ehlen, Patrick and Conrad, Frederick G and Schober, Michael F and Antoun, Christopher and Fail, Stefanie and Hupp, Andrew and Vickers, Lucas and Yan, Huiying and Zhang, Chan},
  booktitle={{Proceedings of the SIGDIAL 2013 Conference}},
  pages={329--333},
  year={2013}
}

@inproceedings{conrad2013mode,
  title={{Mode choice on an iPhone increases survey data quality}},
  author={Conrad, Frederick G and Schober, Michael F and Zhang, Chan and Yan, Huiying and Vickers, Lucas and Johnston, Michael and Hupp, Andrew and Hemingway, Lloyd and Fail, Stefanie and Ehlen, Patrick and others},
  booktitle={{Annual Conference of the American Association for Public Opinion Research}, Boston},
  year={2013}
}

@book{conrad2007envisioning,
  title={Envisioning the Survey Interview of the Future},
  author={Conrad, F.G. and Schober, M.F.},
  isbn={9780470183366},
  series={Wiley Series in Survey Methodology},
  url={https://books.google.co.uk/books?id=fpRP1WJRbz4C},
  year={2007},
  publisher={Wiley}
}

@article{wei2024leveraging,
  title={{Leveraging large language models to power chatbots for collecting user self-reported data}},
  author={Wei, Jing and Kim, Sungdong and Jung, Hyunhoon and Kim, Young-Ho},
  journal={Proceedings of the ACM on Human-Computer Interaction},
  volume={8},
  number={CSCW1},
  pages={1--35},
  year={2024},
  publisher={ACM New York, NY, USA}
}

@inproceedings{devault2014simsensei,
  title={{SimSensei Kiosk: A virtual human interviewer for healthcare decision support}},
  author={DeVault, David and Artstein, Ron and Benn, Grace and Dey, Teresa and Fast, Ed and Gainer, Alesia and Georgila, Kallirroi and Gratch, Jon and Hartholt, Arno and Lhommet, Margaux and others},
  booktitle={{Proceedings of the 2014 international conference on Autonomous agents and multi-agent systems}},
  pages={1061--1068},
  year={2014}
}

@article{inoue2020job,
  title={{A job interview dialogue system that asks follow-up questions: Implementation and evaluation with an autonomous android}},
  author={Inoue, Koji and Hara, Kohei and Lala, Divesh and Nakamura, Shizuka and Takanashi, Katsuya and Tatsuya, Kawahara},
  journal={Transactions of the Japanese Society for Artificial Intelligence},
  volume={35},
  number={5},
  pages={D--K43},
  year={2020}
}

@article{nagasawa2023adaptive,
  title={{Adaptive Interview Strategy Based on Interviewees Speaking Willingness Recognition for Interview Robots}},
  author={Nagasawa, Fuminori and Okada, Shogo and Ishihara, Takuya and Nitta, Katsumi},
  journal={IEEE Transactions on Affective Computing},
  year={2023},
  publisher={IEEE}
}

@article{ge2022should,
  title={{What should i ask: A knowledge-driven approach for follow-up questions generation in conversational surveys}},
  author={Ge, Yubin and Xiao, Ziang and Diesner, Jana and Ji, Heng and Karahalios, Karrie and Sundaram, Hari},
  journal={arXiv preprint arXiv:2205.10977},
  year={2022}
}

@inproceedings{zeng2023question,
  title={{Question Generation to Elicit Users’ Food Preferences by Considering the Semantic Content}},
  author={Zeng, Jie and Nakano, Yukiko I and Sakato, Tatsuya},
  booktitle={{Proceedings of the 24th Annual Meeting of the Special Interest Group on Discourse and Dialogue}},
  pages={190--196},
  year={2023}
}

@article{wuttke2024ai,
  title={{AI Conversational Interviewing: Transforming Surveys with LLMs as Adaptive Interviewers}},
  author={Wuttke, Alexander and A{\ss}enmacher, Matthias and Klamm, Christopher and Lang, Max M and W{\"u}rschinger, Quirin and Kreuter, Frauke},
  journal={arXiv preprint arXiv:2410.01824},
  year={2024}
}

@article{dubey2024llama,
  title={{The llama 3 herd of models}},
  author={Dubey, Abhimanyu and Jauhri, Abhinav and Pandey, Abhinav and Kadian, Abhishek and Al-Dahle, Ahmad and Letman, Aiesha and Mathur, Akhil and Schelten, Alan and Yang, Amy and Fan, Angela and others},
  journal={arXiv preprint arXiv:2407.21783},
  year={2024}
}

@article{jiang2021supporting,
  title={{Supporting serendipity: Opportunities and challenges for Human-AI Collaboration in qualitative analysis}},
  author={Jiang, Jialun Aaron and Wade, Kandrea and Fiesler, Casey and Brubaker, Jed R},
  journal={Proceedings of the ACM on Human-Computer Interaction},
  volume={5},
  number={CSCW1},
  pages={1--23},
  year={2021},
  publisher={ACM New York, NY, USA}
}

@article{bach2024unpacking,
  title={{Unpacking human-AI interaction in safety-critical industries: a systematic literature review}},
  author={Bach, Tita A and Kristiansen, Jenny K and Babic, Aleksandar and Jacovi, Alon},
  journal={IEEE Access},
  year={2024},
  publisher={IEEE}
}

@article{likert1932technique,
  title={{A technique for the measurement of attitudes}},
  author={Likert, Rensis},
  journal={Archives of Psychology},
  year={1932}
}

@misc{reichheld2003the,
	title = {The {One} {Number} {You} {Need} to {Grow}},
	url = {https://hbr.org/2003/12/the-one-number-you-need-to-grow},
	urldate = {2025-02-18},
        journal ={Harvard Business Review},
        year={2003},
        author={Reichheld, Fred},
}

@article{flesch1948new,
  title={{A new readability yardstick.}},
  author={Flesch, Rudolph},
  journal={Journal of applied psychology},
  volume={32},
  number={3},
  pages={221},
  year={1948},
  publisher={American Psychological Association}
}

@misc{cuevas_collecting_2024,
	title = {Collecting {Qualitative} {Data} at {Scale} with {Large} {Language} {Models}: {A} {Case} {Study}},
	shorttitle = {Collecting {Qualitative} {Data} at {Scale} with {Large} {Language} {Models}},
	url = {http://arxiv.org/abs/2309.10187},
	doi = {10.48550/arXiv.2309.10187},
	urldate = {2025-02-19},
	publisher = {arXiv},
	author = {Cuevas, Alejandro and Scurrell, Jennifer V. and Brown, Eva M. and Entenmann, Jason and Daepp, Madeleine I. G.},
	month = dec,
	year = {2024},
	note = {arXiv:2309.10187 [cs]},
}

@misc{uberti_webrtc_2024,
	title = {{WebRTC}: {Real}-{Time} {Communication} in {Browsers}},
	url = {https://www.w3.org/TR/webrtc/},
	urldate = {2025-02-19},
	publisher = {Google},
	author = {Uberti, Justin and Thatcher, Peter and Brandstetter, Taylor and Aboba, Bernard and Narayanan, Anant and Bergkvist, Adam and Burnett, Daniel C.},
	year = {2024},
}

@article{definitions2011final,
  title={{Final dispositions of case codes and outcome rates for surveys}},
  author={AAPOR},
  journal={The American Association for Public Opinion Research},
  year={2023}
}

@article{abdulhai2023moral,
  title={{Moral foundations of large language models}},
  author={Abdulhai, Marwa and Serapio-Garcia, Gregory and Crepy, Cl{\'e}ment and Valter, Daria and Canny, John and Jaques, Natasha},
  journal={arXiv preprint arXiv:2310.15337},
  year={2023}
}

@article{ranaldi2023large,
  title={{When Large Language Models contradict humans? Large Language Models' Sycophantic Behaviour}},
  author={Ranaldi, Leonardo and Pucci, Giulia},
  journal={arXiv preprint arXiv:2311.09410},
  year={2023}
}

@article{rottger2023xstest,
  title={{Xstest: A test suite for identifying exaggerated safety behaviours in large language models}},
  author={R{\"o}ttger, Paul and Kirk, Hannah Rose and Vidgen, Bertie and Attanasio, Giuseppe and Bianchi, Federico and Hovy, Dirk},
  journal={arXiv preprint arXiv:2308.01263},
  year={2023}
}

@inproceedings{nikolenko2023attitude,
  title={{The attitude of young people to the use of artificial intelligence}},
  author={Nikolenko, Oksana and Astapenko, Evgeniya},
  booktitle={{E3S Web of Conferences}},
  volume={460},
  pages={05013},
  year={2023},
  organization={EDP Sciences}
}

@article{horowitz2024adopting,
  title={{Adopting AI: how familiarity breeds both trust and contempt}},
  author={Horowitz, Michael C and Kahn, Lauren and Macdonald, Julia and Schneider, Jacquelyn},
  journal={AI \& society},
  volume={39},
  number={4},
  pages={1721--1735},
  year={2024},
  publisher={Springer}
}

@article{irvine2013not,
  title={{‘Am I not answering your questions properly?’Clarification}, adequacy and responsiveness in semi-structured telephone and face-to-face interviews},
  author={Irvine, Annie and Drew, Paul and Sainsbury, Roy},
  journal={Qualitative research},
  volume={13},
  number={1},
  pages={87--106},
  year={2013},
  publisher={Sage Publications Sage UK: London, England}
}

@article{levin1998automatic,
  title={{Automatic evaluation of spoken dialogue systems}},
  author={Levin, Esther},
  year={1998}
}

@inproceedings{zeigler1994dialog,
  title={{Dialog design for a speech-interactive automation system}},
  author={Zeigler, BL and Bazor, B},
  booktitle={{Proceedings of 2nd IEEE Workshop on Interactive Voice Technology for Telecommunications Applications}},
  pages={113--116},
  year={1994},
  organization={IEEE}
}

@article{cole1997experiments,
  title={{Experiments with a spoken dialogue system for taking the US census}},
  author={Cole, Ronald A and Novick, David G and Vermeulen, Pieter JE and Sutton, Stephen and Fanty, Mark and Wessels, LFA and de Villiers, Jacques Ho and Schalkwyk, Johan and Hansen, Brian and Burnett, Daniel},
  journal={Speech Communication},
  volume={23},
  number={3},
  pages={243--260},
  year={1997},
  publisher={Elsevier}
}

@article{singh1999reinforcement,
  title={{Reinforcement learning for spoken dialogue systems}},
  author={Singh, Satinder and Kearns, Michael and Litman, Diane and Walker, Marilyn},
  journal={Advances in neural information processing systems},
  volume={12},
  year={1999}
}

@article{boyce2000natural,
  title={{Natural spoken dialogue systems for telephony applications}},
  author={Boyce, Susan J},
  journal={Communications of the ACM},
  volume={43},
  number={9},
  pages={29--34},
  year={2000},
  publisher={ACM New York, NY, USA}
}

@article{groves2008telephone,
  title={{Telephone interviewer voice characteristics and the survey participation decision}},
  author={Groves, Robert M and O’Hare, Barbara C and Gould-Smith, Dottye and Benk{\'\i}, Jos{\'e} and Maher, Patty and Hansen, SE},
  journal={Advances in telephone survey methodology},
  pages={385--400},
  year={2008},
  publisher={Wiley Online Library}
}

@inproceedings{lin2022duplex,
  title={{Duplex conversation: Towards human-like interaction in spoken dialogue systems}},
  author={Lin, Ting-En and Wu, Yuchuan and Huang, Fei and Si, Luo and Sun, Jian and Li, Yongbin},
  booktitle={{Proceedings of the 28th ACM SIGKDD Conference on Knowledge Discovery and Data Mining}},
  pages={3299--3308},
  year={2022}
}

@inproceedings{jin2021duplex,
  title={{Duplex Conversation in Outbound Agent System.}},
  author={Jin, Chunxiang and Yang, Minghui and Wen, Zujie},
  booktitle={{Interspeech}},
  pages={4866--4867},
  year={2021}
}

@article{lu2025duplexmamba,
  title={{DuplexMamba: Enhancing Real-time Speech Conversations with Duplex and Streaming Capabilities}},
  author={Lu, Xiangyu and Xu, Wang and Wang, Haoyu and Zhou, Hongyun and Zhao, Haiyan and Zhu, Conghui and Zhao, Tiejun and Yang, Muyun},
  journal={arXiv preprint arXiv:2502.11123},
  year={2025}
}

@article{wang2024full,
  title={{A full-duplex speech dialogue scheme based on large language models}},
  author={Wang, Peng and Lu, Songshuo and Tang, Yaohua and Yan, Sijie and Xia, Wei and Xiong, Yuanjun},
  journal={arXiv preprint arXiv:2405.19487},
  year={2024}
}

@article{zhang2025llm,
  title={{LLM-Enhanced Dialogue Management for Full-Duplex Spoken Dialogue Systems}},
  author={Zhang, Hao and Li, Weiwei and Chen, Rilin and Kothapally, Vinay and Yu, Meng and Yu, Dong},
  journal={arXiv preprint arXiv:2502.14145},
  year={2025}
}

@article{schober2015precision,
  title={{Precision and disclosure in text and voice interviews on smartphones}},
  author={Schober, Michael F and Conrad, Frederick G and Antoun, Christopher and Ehlen, Patrick and Fail, Stefanie and Hupp, Andrew L and Johnston, Michael and Vickers, Lucas and Yan, H Yanna and Zhang, Chan},
  journal={PloS one},
  volume={10},
  number={6},
  pages={e0128337},
  year={2015},
  publisher={Public Library of Science San Francisco, CA USA}
}

@article{van2004event,
  title={{Event-related potentials indicate motivational relevance of cocaine cues in abstinent cocaine addicts}},
  author={van der Laar, MC and Licht, R and Franken, HA and Hendriks, VM},
  journal={Psychopharmacology},
  volume={177},
  number={1/2},
  pages={121--129},
  year={2004},
  publisher={Springer Verlag}
}

@article{wang2024telephone,
  title={{Telephone follow-up based on artificial intelligence technology among hypertension patients: Reliability study}},
  author={Wang, Siyuan and Shi, Yan and Sui, Mengyun and Shen, Jing and Chen, Chen and Zhang, Lin and Zhang, Xin and Ren, Dongsheng and Wang, Yuheng and Yang, Qinping and others},
  journal={The Journal of Clinical Hypertension},
  year={2024},
  publisher={Wiley Online Library}
}

@article{hong2022,
author={Hong,Grace and Smith,Margaret and Lin,Steven},
year={2022},
month={06},
title={The AI Will See You Now: Feasibility and Acceptability of a Conversational AI Medical Interviewing System},
journal={JMIR Formative Research},
volume={6},
number={6},
note={Copyright - © 2022. This work is licensed under https://creativecommons.org/licenses/by/4.0/ (the “License”). Notwithstanding the ProQuest Terms and Conditions, you may use this content in accordance with the terms of the License; Last updated - 2023-11-26},
language={English},
url={https://www.proquest.com/scholarly-journals/ai-will-see-you-now-feasibility-acceptability/docview/2682567709/se-2},
}

\newpage

\begin{appendices}
\renewcommand{\thefigure}{A\arabic{figure}}
\setcounter{figure}{0} 

\section{Appendix}
\begin{figure}[htbp]
    \centering
    \includegraphics[width=0.65\linewidth]{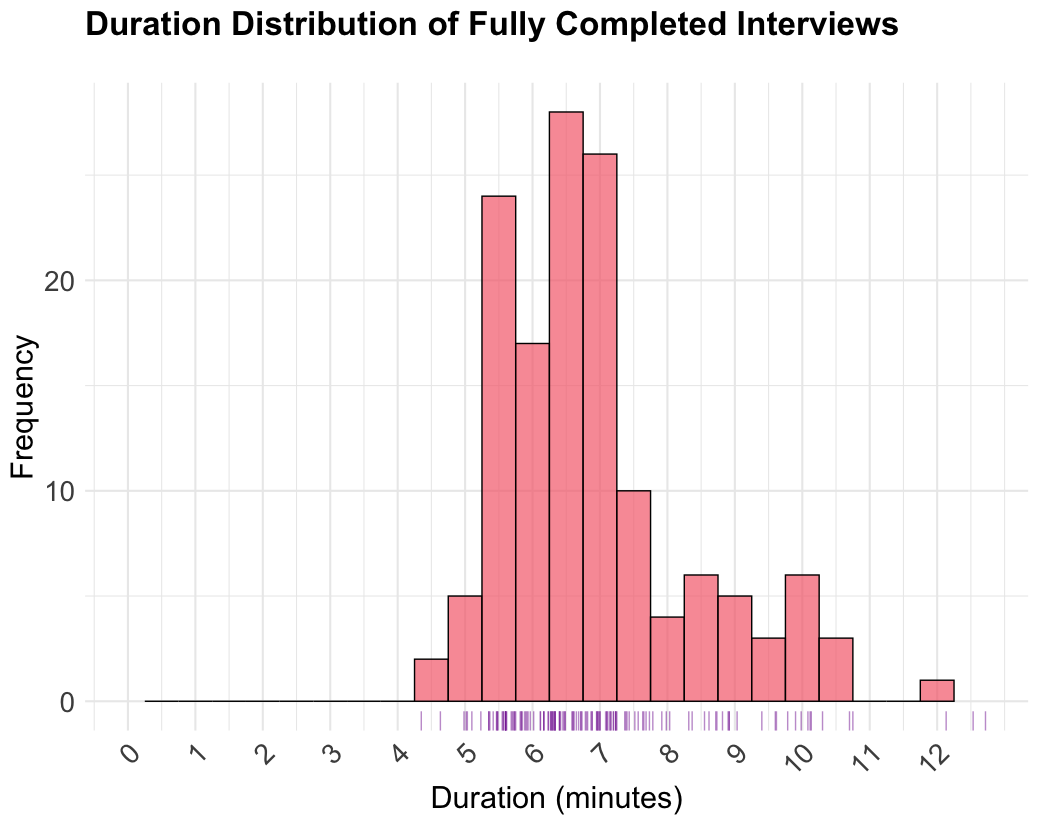}
    \caption{Distribution of fully completed interview durations (in minutes).}
    \label{fig:call_duration}
\end{figure}

\end{appendices}

\end{document}